\documentclass[epj-spec]{svjour}
\usepackage{graphics}
\usepackage{amssymb}
\usepackage{amsmath}
\usepackage{amsfonts}
\usepackage{epsfig}
\usepackage{hyperref}
\begin{document}
\title{Increasing Average Period Lengths by Switching of Robust Chaos Maps in Finite Precision}
\author{Nithin Nagaraj\inst{1}\fnmsep\thanks{\email{nithin\_nagaraj@yahoo.com},~\href{http://nithin.nagaraj.googlepages.com}{http://nithin.nagaraj.googlepages.com}} \and Mahesh C. Shastry\inst{2}\fnmsep\thanks{\email{mcs312@psu.edu},~\href{http://www.personal.psu.edu/mcs312}{http://www.personal.psu.edu/mcs312}} \and Prabhakar G. Vaidya\inst{1}}
\institute{School of Natural Sciences and Engineering, National
Institute of Advanced Studies, IISc. Campus, Bangalore 560 012,
INDIA \and Electrical Engineering Department, The Pennsylvania State
University, PA 16801, USA}
\abstract{
Grebogi, Ott and Yorke (Phys. Rev. A 38(7), 1988) have investigated
the effect of finite precision on average period length of chaotic
maps. They showed that the average length of periodic orbits ($T$)
of a dynamical system scales as a function of computer precision
($\varepsilon$) and the correlation dimension ($d$) of the chaotic
attractor: $T \sim \varepsilon^{-d/2}$. In this work, we are
concerned with increasing the average period length which is
desirable for chaotic cryptography applications. Our experiments
reveal that random and chaotic switching of deterministic chaotic
dynamical systems yield higher average length of periodic orbits as
compared to simple sequential switching or absence of switching. To
illustrate the application of switching, a novel generalization of
the Logistic map that exhibits Robust Chaos (absence of attracting
periodic orbits) is first introduced. We then propose a
pseudo-random number generator based on chaotic switching between
Robust Chaos maps which is found to successfully pass stringent
statistical tests of randomness.} 
\maketitle
\section{Introduction}
\label{sec:1} 
\begin{figure}[!h]
\centering
\includegraphics[scale=.3]{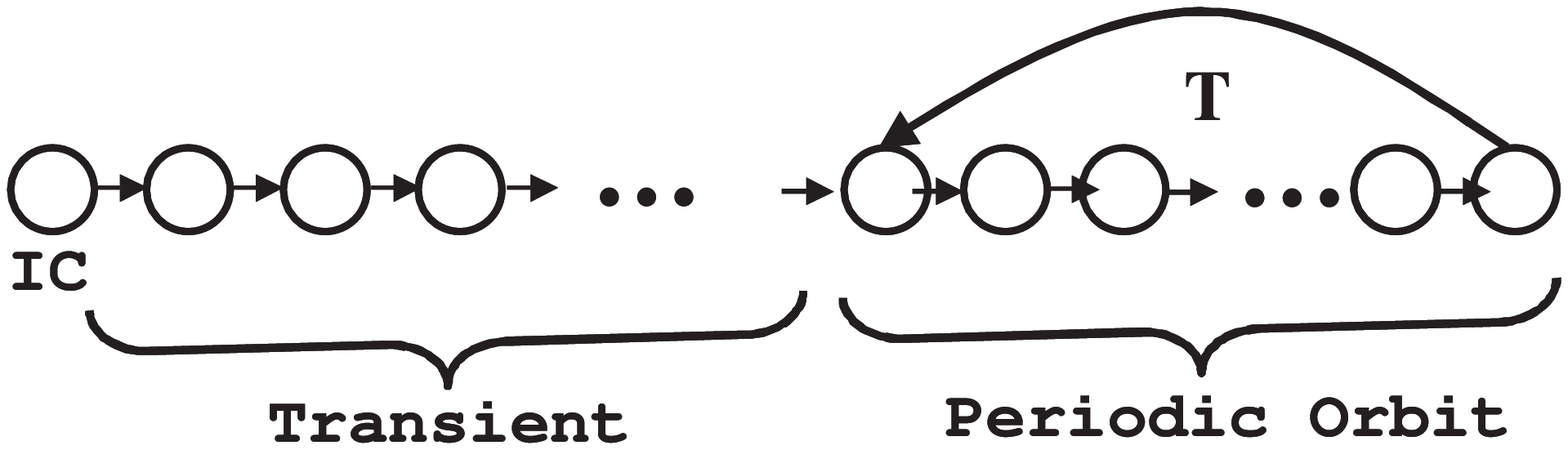}
\hspace{0.1in}
\includegraphics[scale=.4]{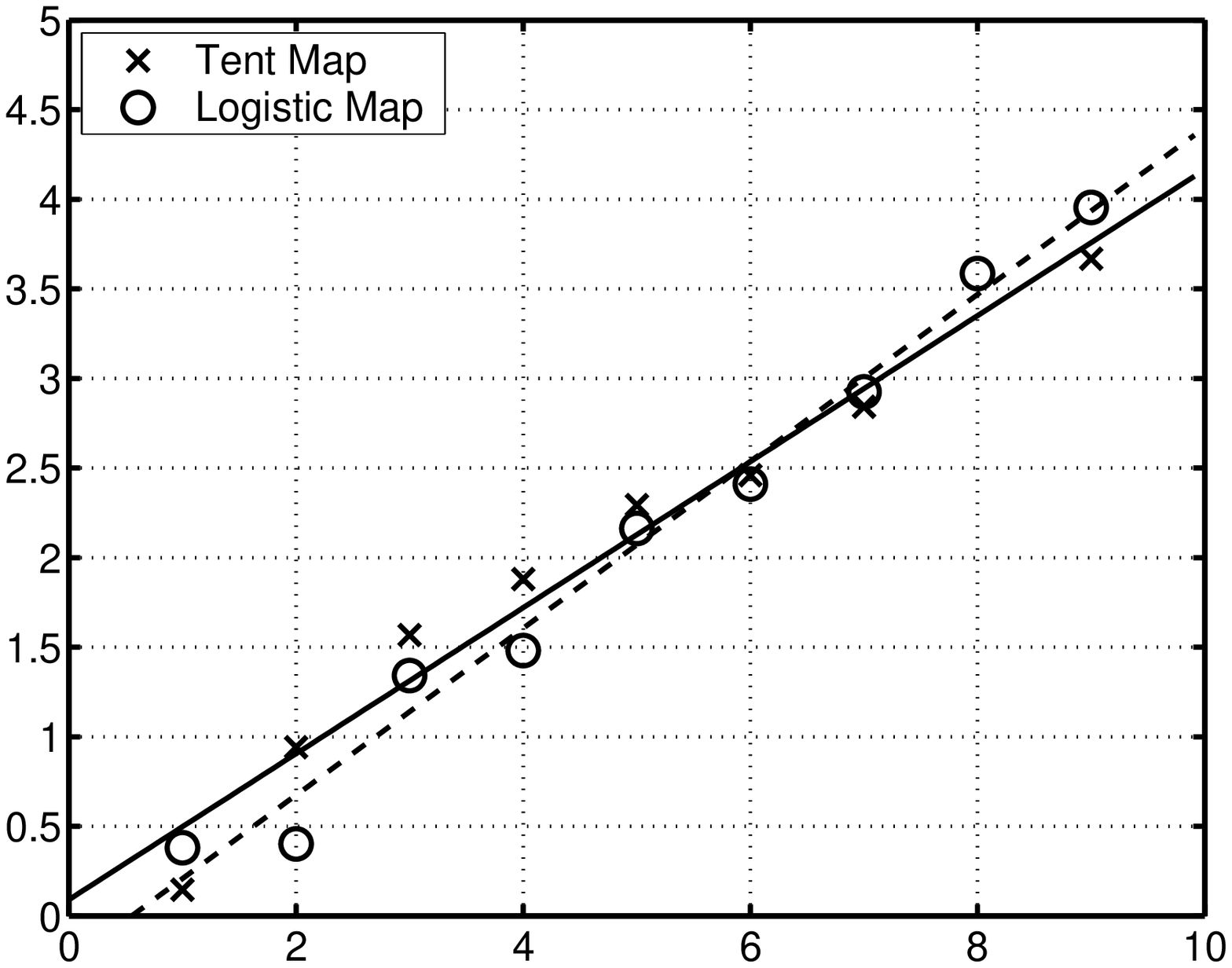}
 \caption{(a) Left: All initial conditions of a
dynamical system when iterated on a finite precision computer end up
in a periodic orbit. (b) Right: Average period length $T$ vs.
$(1/\varepsilon)$ in logarithmic scale for the Tent map ($T =
10^{0.0904} \varepsilon^{-0.4075}$) and the Logistic map ($T =
10^{-0.2576} \varepsilon^{-0.4658}$). This agrees quite well with
the relationship derived by Grebogi~\cite{Grebogi}.}
\label{fig:figchaoscomp}
\end{figure}

Grebogi, Ott and Yorke (Phys. Rev. A 38(7), 1988) were
one of the first to study in detail the effect of finite precision
on the expected length of periodic orbits and their distribution for
chaotic maps. On a digital computer, since there are only a finite
number of states owing to limited precision, all autonomous chaotic
maps (and flows) when simulated would have to eventually settle down
to periodic orbits for all initial conditions
(Figure~\ref{fig:figchaoscomp}(a), $T$ is defined as the period
length). Thus Chaos which is characterized by the existence of
wandering orbits which have infinite period length is impossible on
a digital computer.

\par This fact is often under appreciated.  Chaotic cryptographic
applications appeal to the fact that Chaos has inherently good
mixing properties which are suitable for confusing and diffusing the
message. However, owing to limited precision, this is not strictly
true since there would be only a finite number of periodic orbits
and no wandering orbits. This probably explains why many chaotic
cryptographic algorithms have been eventually broken though they
initially promised to exhibit strong theoretical security.

The question that naturally arises is: What is the average length of
the periodic orbits of the dynamical system implemented on a finite
precision computer? Grebogi's work~\cite{Grebogi} showed that the
average length of periodic orbits ($T$) of a dynamical system scales
as a function of computer precision ($\varepsilon$) and the
correlation dimension ($d$, it is defined as the exponent of the
power-law dependence of the correlation integral and is a measure of
the strangeness of the attractor~\cite{Grassberger}) of the chaotic
attractor: $T \sim \varepsilon^{-d/2}$.
Figure~\ref{fig:figchaoscomp}(b) shows the plot of the average
period length $T$ with respect to $(1/\varepsilon)$ (logarithmic
scale) for the Tent map ($x \mapsto 2x, 0 \leq x < 0.5;~~ x \mapsto
2-2x, 0.5 \leq x \leq 1$) and the Logistic map ($x \mapsto
4x(1-x)$). In both cases, $x \in [0,1]$. As it can be seen, the
relationship $T \sim \varepsilon^{-d/2}$ is empirically well matched
in both cases.

Subsequently, there have been only few studies on the effect of
computer precision on the average period length. Wang et.
al.~\cite{Wang} discovered that a single dominant periodic
trajectory is realized with major probability. They also studied
coupled maps and showed that these exhibit larger period lengths
which can be useful in chaotic cryptography. The other notable work
with regards to the effect of computer precision to chaotic
cryptography is that of Li~\cite{Li} where they performed
quantitative analysis of the degradation of digitized chaos and
proposed a new series of dynamical indicators for 1D piecewise
linear chaotic maps.

The question that we are interested in is -- How can we increase the
average period length $T$ while still using the same precision? This
paper is organized as follows. In Section 2, the effect of switching
between chaotic maps on the average period length and how different
switching strategies increase the average period length is
discussed. In Section 3, a pseudo-random number generator is built
to demonstrate the applications of switching of chaotic maps. To
this end, a generalization of the Logistic map that exhibits Robust
Chaos is proposed. Robust Chaos is theoretically characterized by
absence of attracting periodic orbits. This property is highly
desirable for chaotic cryptographic applications. The pseudo-random
number generator switches between these family of maps (still under
finite precision). In Section 4, randomness evaluation of the
pseudo-random number generator to demonstrate its merit is performed
and we conclude in Section 5.
\section{Switching of chaotic maps and its effect on average period length}
\label{sectionswitching} We consider switching between chaotic maps
and study the effect of different switching strategies on the
average period length. What we mean by switching of maps is as
follows. We start with an initial condition (typically chosen at
random) and iterate with the first dynamical system (say the Tent
map). For the sequential switching strategy, after one iteration of
the first dynamical system, we iterate the second dynamical system
(say the Logistic map). We then iterate the first dynamical system
again and so on and so forth. For a chaotic or random switching
strategy, we randomly or chaotically choose the dynamical system to
iterate at every iteration. We then determine the period length and
repeat this for a number of initial conditions and take their
average. We do this for several precision values.

\begin{figure}[!h]
\centering
\includegraphics[scale=.38]{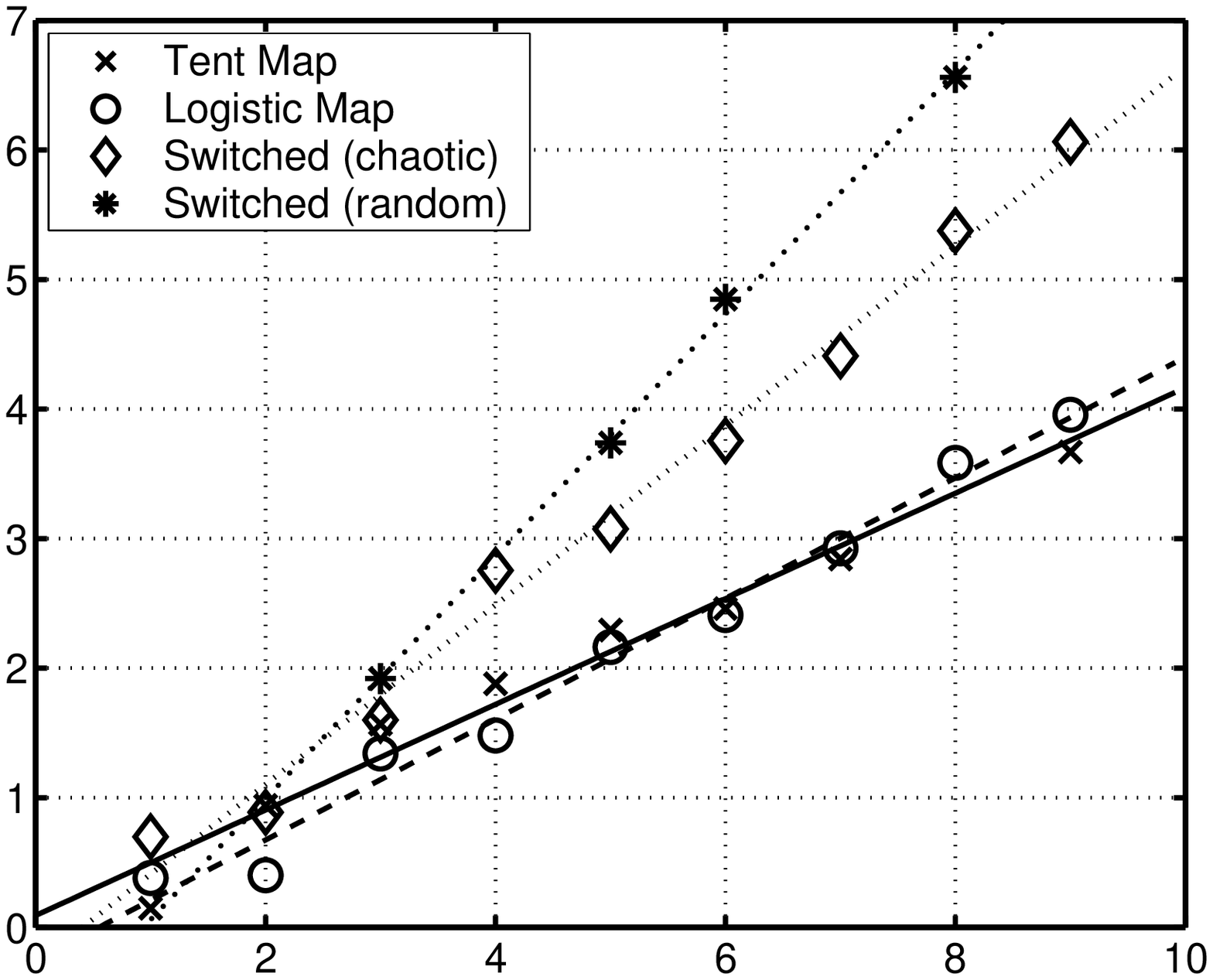}
\hspace{0.01in}
\includegraphics[scale=.35]{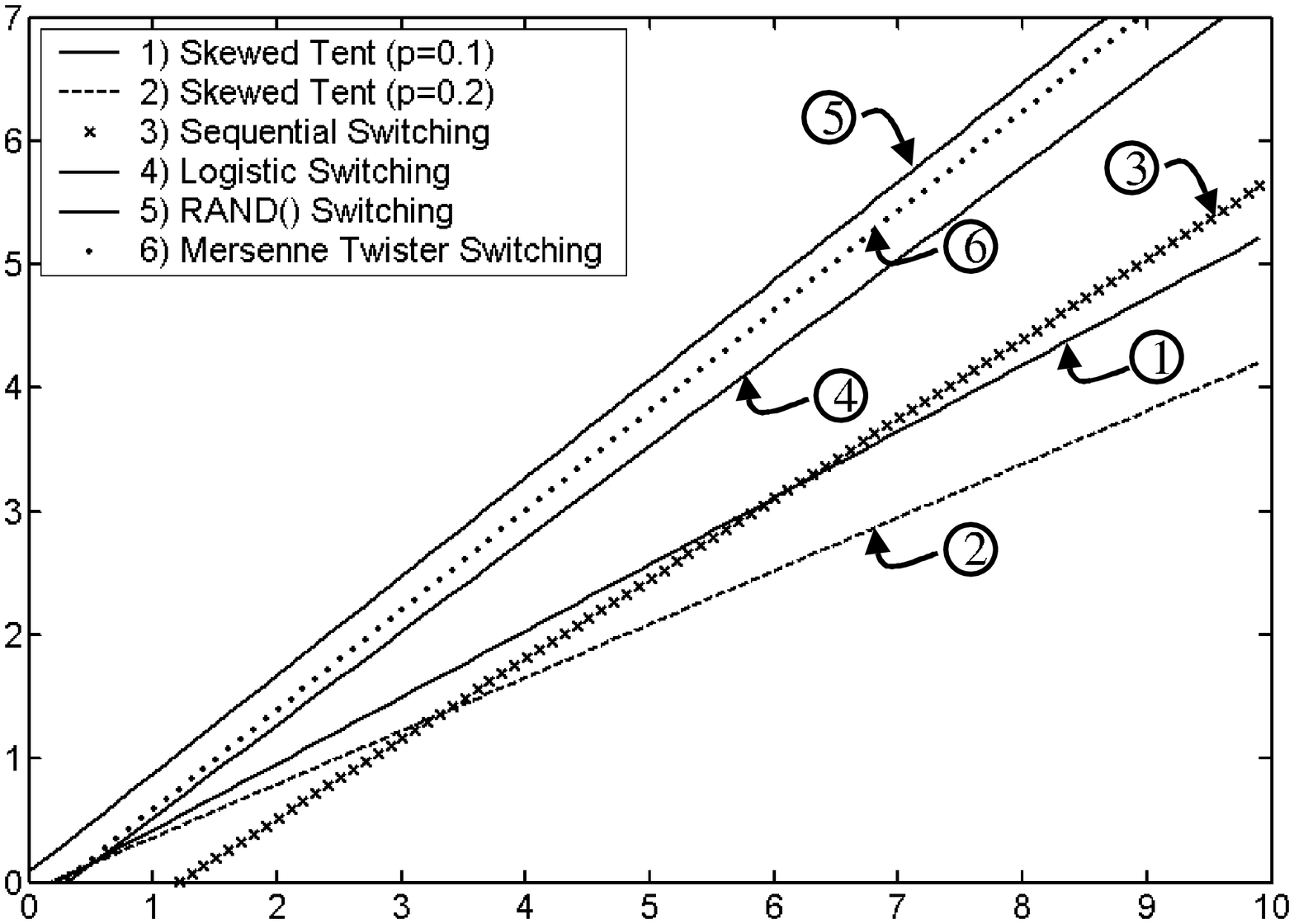}
 \caption{Effect of switching on average period length: (a) Left: Chaotic switching yields $T =
10^{-0.2824} \varepsilon^{-0.6924}$ and Random switching using
RAND(.) yields $T = 10^{-0.8766} \varepsilon^{-0.9351}$. Sequential
switching (not shown in graph) yields $T = 10^{0.0137}
\varepsilon^{-0.4270}$. (b) Right: Similar results were obtained for
switching between skewed Tent maps (p=0.1 and 0.2). Observe that in
both cases, $T$ follows the following order: No switching $<$
Sequential switching $<$ Chaotic switching $<$ Random switching.}
\label{fig:figswitch}
\end{figure}

We chose the Tent map, the Logistic map and the Skewed Tent map ($x
\mapsto \frac{x}{p}, 0 \leq x < p;~~ x \mapsto \frac{1-x}{1-p},
p\leq x \leq 1$) with values of $p=0.1$ and $p=0.2$. The switching
strategies we used were sequential, chaotic (Logistic map) and
random (Mersenne Twister~\cite{MersenneTwister} and RAND(.)). The
experiments were performed on a computer with the following
specifications -- Pentium$^{\circledR}$ 4, CPU 3.06 GHz, 480 M bytes
RAM (C program). For smaller $\varepsilon$, we averaged the period
lengths over 1000 randomly chosen initial conditions. The results
are shown in Figure~\ref{fig:figswitch}.

It can be inferred from Figure~\ref{fig:figswitch} that switching
increases the average period length as indicated by the increase in
the slope of the linear fit.

\section{Pseudo-random number generator using switching}
 It is well known that chaotic dynamical
systems exhibit unpredictability, ergodicity and mixing properties.
This suggests that chaotic maps can be used in generating
pseudo-random numbers. Pseudo-random numbers are those which are
``random-like'' in their statistical properties.  In 1947, Ulam and
von Neumann suggested using the Logistic map to generate a sequence
of pseudo-random numbers. Pseudo-random numbers are used in a
variety of applications such as in Monte-Carlo simulations for
random sampling from a distribution and are central in cryptographic
applications to build stream and block ciphers and in several
protocols requiring generation of random data. Pseudo-random number
generators are algorithms implemented on digital systems that can
generate these numbers. Due to limitations in computation and
precision, pseudo-random number sequences are necessarily periodic
(as opposed to an ideal random number generator which is a discrete
memoryless information source that generates equiprobable
non-periodic symbols; we shall not discuss these here). Sequences
generated by pseudo-random number generators are expected to have
large periods and pass a number of statistical randomness tests. In
this paper, the phrase \emph{random numbers} refers to uniformly
distributed pseudo-random numbers.

The relationship between chaos and cryptography has been discussed
by Kocarev~\cite{kocarev 1}. Various one-dimensional chaotic maps
have been proposed for generating random numbers, e.g:
PL1D~\cite{kocarev 2}, LOGMAP~\cite{logmap} etc. In their study of
the Logistic Map, Pathak and Rao~\cite{logmap} propose a
pseudo-random number generator which has a period of about $10^8$
when implemented in double precision. This period is quite small
when compared to many other `good' random number generators in the
literature. They conjecture that such a period is due to the fact
that the value of $a$ in the Logistic map: $ax(1-x)$, becomes
slightly less than $4$  because of which the map settles down into
windows (attracting periodic orbits, see
Figure~\ref{fig:figfragilerobustbif}(a)). We define the absence of
attracting periodic orbits (for a particular value of the
bifurcation parameter) as `Full Chaos'. For the Logistic map, $a=4$
exhibits `Full Chaos'. The fact that Full Chaos exists only for a
small set of parameters is a major hindrance in using chaotic maps
as pseudo-random number generators. The limitations of computation
and precision cause the parameters to deviate from Full Chaos values
and this may result in settling on an attracting periodic orbit.
Another major disadvantage of using chaotic maps directly is that
the the successive points are strongly correlated. This shows up in
the 2-dimensional phase space plot of the iterates.

To overcome the problem of low period lengths when using chaotic
maps on finite precision computer, one of the strategies might be to
switch between different chaotic maps and use them as a sequence of
random numbers. We have already shown in
Section~\ref{sectionswitching} that switching between chaotic maps
increases the average period length. We have also found that
switching between more number of maps increases the average period
length further (we have not indicated these results for want of
space). However, we still need to evolve a strategy to avoid
settling into windows or attracting periodic orbits (only under
infinite precision, in finite precision it is still an open problem
whether this has any benefits or not). To this end, we suggest using
maps with the special property that they exhibit Full Chaos for a
neighborhood of the parameter space unlike the logistic map which
exhibits Full Chaos only at a single point ($a=4$). This special
property is defined as Robust Chaos. Chaos which fails to satisfy
this special property is termed as Fragile Chaos (for eg., Logistic
map, Tent map and most well known maps exhibit Fragile Chaos).

%
%
\subsection{Robust Chaos maps}

Robust Chaos is defined by the absence of periodic windows and
coexisting attractors in some neighborhood of the parameter
space~\cite{Banerjee}. Barreto~\cite{Barreto} had conjectured that
robust chaos may not be possible in smooth unimodal one-dimensional
maps. This was shown to be false with counter-examples by
Andrecut~\cite{Andrecut1} and Banerjee~\cite{Banerjee}. Banerjee
demonstrates the use of robust chaos in a practical example in
electrical engineering. Andrecut provides a general procedure for
generating robust chaos in smooth unimodal maps~\cite{Andrecut2}.

As observed by Andrecut~\cite{Andrecut1}, robust chaos implies a
kind of ergodicity or good mixing properties of the map. This makes
it very beneficial for cryptographic purposes. The absence of
windows would mean that the these maps can be used in hardware
(analog) implementation as there would be no {\it fragility} of
chaos with noise (for eg., thermal noise) induced variation of the
parameters (it would be impossible in practice to maintain a
constant value, $a=4$, in any analog implementation of Logistic
Map). As it is impossible to eliminate noise in any hardware
(analog) implementation, this property of Robust Chaos would be
beneficial.


\subsubsection{Some maps where we encountered Robust Chaos}
A list of maps that we found to exhibit Robust Chaos are given
below. Here $x \in [0,1]$.
\begin{enumerate}
\item $x \mapsto \beta x - \lfloor \beta x \rfloor$ where $\beta$ is any positive real number ($\neq
1$).
\item $x \mapsto \frac{x}{p}$ if $0 \leq x \leq p$ and $x \mapsto
\frac{1-x}{1-p}$ if $p < x \leq 1$. The well known Tent map belongs
to this family of maps ($p=0.5$).
\item Two parameter Robust Chaos map~\cite{Nithin}: $Skew-nGLS(a,p,x) =
 \frac{(a-p)+\sqrt{(p-a)^2+4ax}}{2a}
   , 0\leq x < p$; $= \frac{(1+a-p)-\sqrt{(p-a-1)^2+4a(1-x)} }{2a},~~~ p \leq x \leq 1,~~~0 < a \leq min(p,1-p).$
\item Andrecut and Ali~\cite{Andrecut2} provide a novel method of converting any chaotic
(not robust) 1D unimodal map $\phi(x)$ (a map that is
$\mathcal{C}^3$ on [0,1] and which contains a single unique critical
point `$c$', actually a maximum) which has negative Schwarzian
derivative to another unimodal map $f_{\nu}^{(\pm)}(x)$ that
exhibits robust chaos given by:
\begin{equation}
f_{\nu}^{(\pm)}(x) = \frac{1 - \nu^{\pm \phi(x)}}{1 - \nu^{\pm
\phi(c)}},~~~ \forall \nu > 0, \nu \neq 1.
\end{equation}
As an example, the following maps, both derived from the unimodal
map $\phi(x) = x(1-x)$ generate robust chaos:
\begin{equation}
f_{\nu}^{+}(x) = \frac{1 - \nu^{+ x(1-x)}}{1 - \nu^{+ 0.25}}, \nu
\in (0,1).~~~~f_{\nu}^{-}(x) = \frac{1 - \nu^{- x(1-x)}}{1 - \nu^{-
0.25}}, \nu \in (1,\infty).
\end{equation}
\item The B-exponential map~\cite{Mahesh} which is a generalization of the Logistic
map exhibits robust chaos. This will be discussed
in~\ref{subsectionBexp}.
\item
One can use the fact that topological conjugacy preserves dynamics
to generate a number of family of maps that all exhibit robust
chaos. As an example, by applying the diffeomorphism $C(x) =
\frac{(1-cos \pi x)}{2}$ to the B-exponential map, we can easily
obtain a generalization of the Tent map that exhibits robust
chaos~\cite{Mahesh}.
\end{enumerate}

\subsection{B-Exponential map $GL(B,x)$}
\label{subsectionBexp} The B-Exponential Map $GL(B,x)$ is defined as
follows:

\begin{equation}
GL(B,x)=\frac{B-xB^x-(1-x)B^{1-x} } {B-\sqrt{B}}, \label{eq:gl}
\qquad 0\leq x \leq 1 \textrm{~and~} B \in \mathbb{R}^+, B \neq 1.
\end{equation}
\\Here, $B$ is the bifurcation parameter. Note that
$x_{n+1}=GL(B,x_n)$ is the iteration function.

\subsubsection{Properties of B-exponential map}

\begin{enumerate}
\item $GL(B,x)$ is unimodal
for $e^{-4} \leq B < \infty$ (unimodal implies that the map has only
one critical point in [0,1] and it passes through zero at 0 and 1).

\item The B-Exponential Map is a generalization of the Logistic map
because of the following interesting property:
\begin{equation}
\lim_{B \rightarrow 1} GL(B,x)=4x(1-x). \label{eq:g2}
\end{equation}
\\This can be derived by a simple application of L'Hospital's
rule. This property is the reason behind the notation `$GL(B,x)$'
where `GL' stands for Generalized Logistic. The maps of $GL(B,x)$
looks similar to the Logistic map for values of $B$ near 1. It is
interesting to note that $GL(B,x)$ tends to a constant function
(with value 1) as B tends to $\infty$ (for all $x$).

\item The Lyapunov exponent of $GL(B,x)$ is a
constant for all $B \geq e^{-4} $ and equals $\ln{2}$. Thus the
B-Exponential map is chaotic for all real $B \geq e^{-4}$. See
~\cite{Mahesh} for details.

\item The B-Exponential Map exhibits {\it Robust Chaos} for $B \geq
e^{-4}$. \label{thm:thm2} See~\cite{Mahesh} for details.


\begin{figure}[!h]
\centering
\includegraphics[scale=.45]{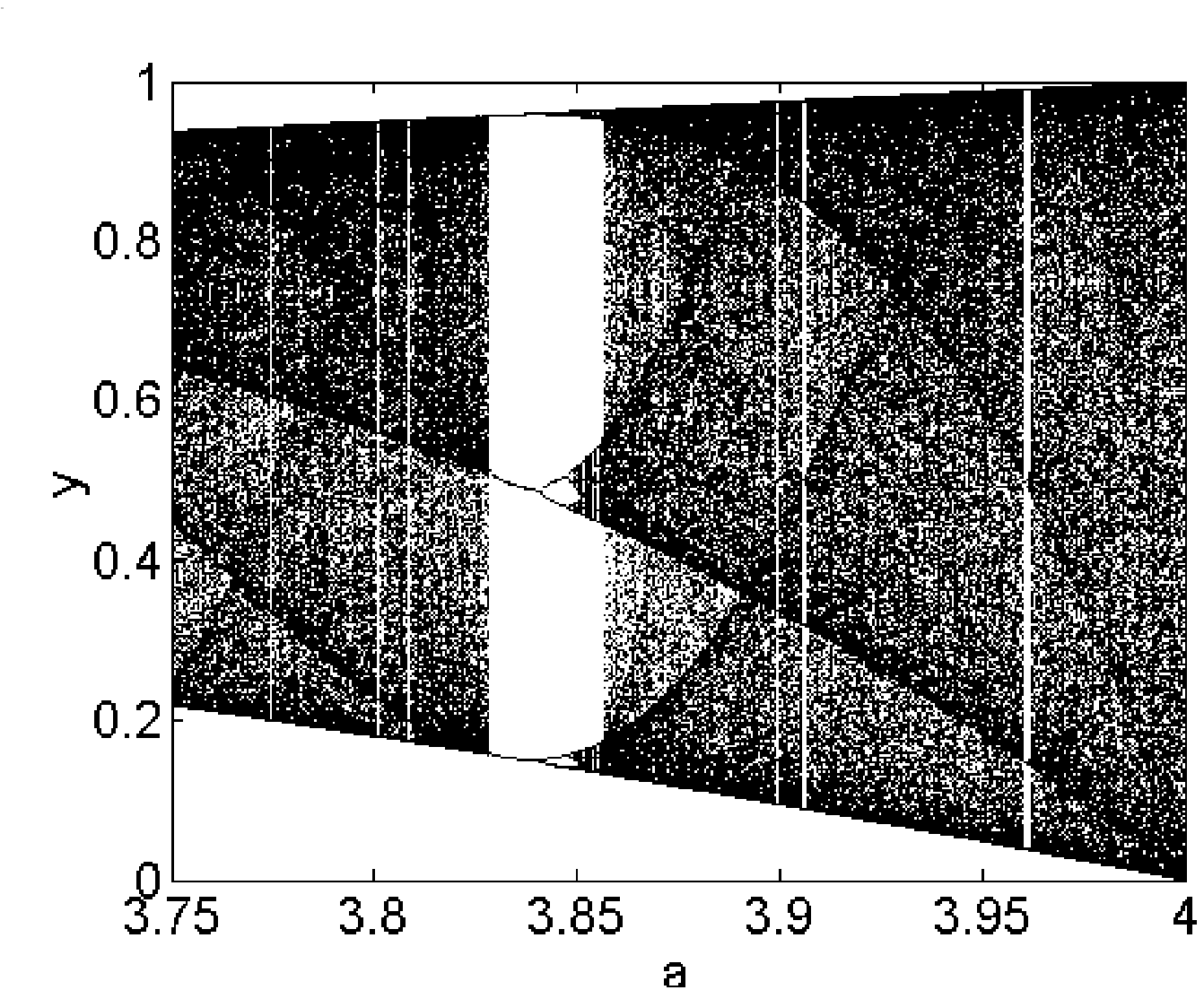}
\hspace{0.01in}
\includegraphics[scale=.45]{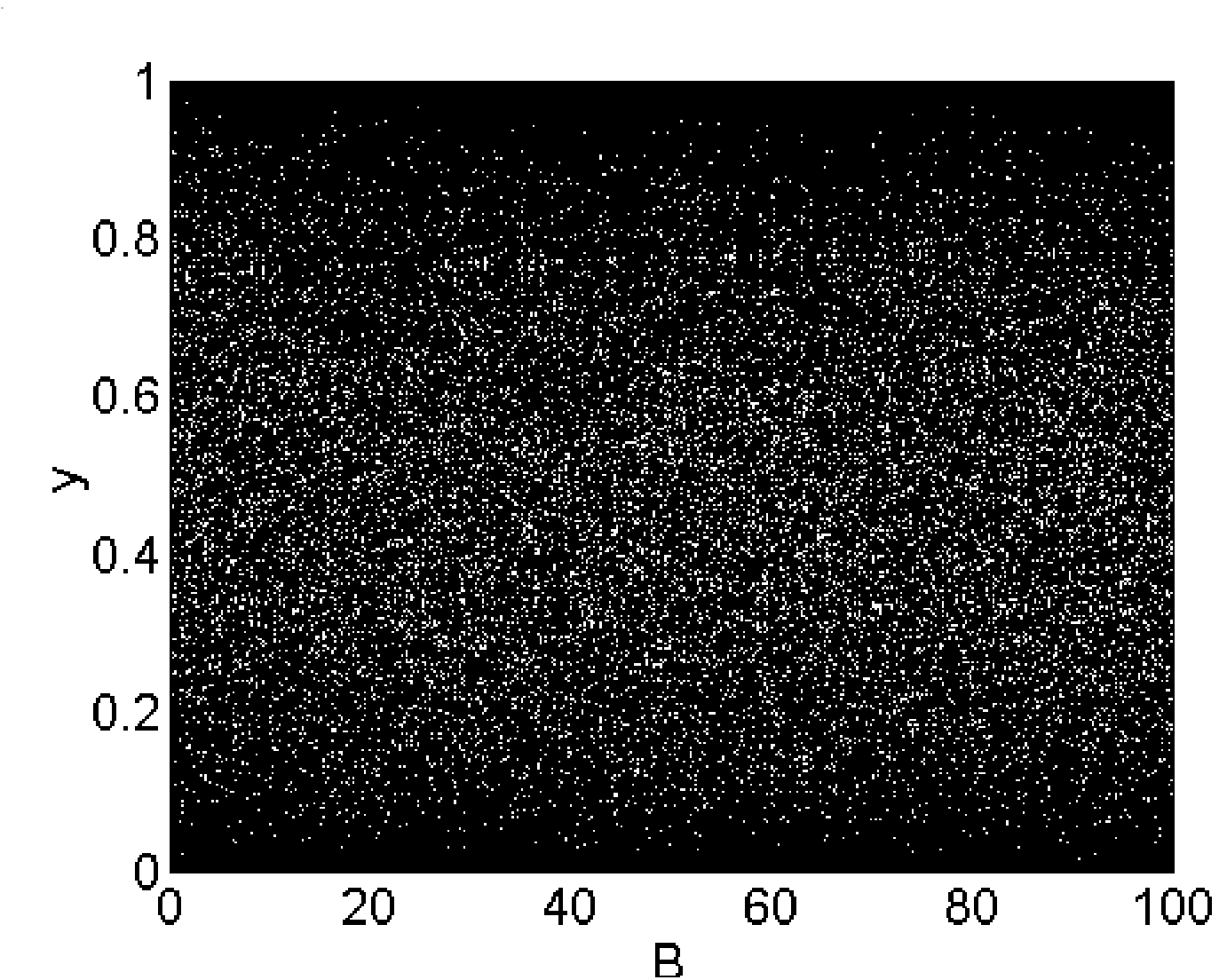}
 \caption{(a) Left: A portion of the bifurcation diagram for the Logistic family ($x \mapsto ax(1-x)$)
 showing attracting periodic orbits (windows). This is termed as
 {\it Fragile} Chaos. (b) Right: Bifurcation diagram for the
 B-exponential map showing no windows. This is termed as {\it Robust} Chaos. This property is useful in generating
 pseudo-random numbers as we shall demonstrate.
 } \label{fig:figfragilerobustbif}
\end{figure}
\end{enumerate}
\subsection{BEACH}
We propose a pseudo-random number generator based on B-Exponential
Map with the name BEACH ({\bf B}-{\bf E}xponential {\bf A}ll-{\bf
C}haotic map-switc{\bf H}ing). As the name suggests, the
pseudo-random number generator is based on the principle of
switching from map to map to extract numbers for the generator. Such
a scheme has been studied by Rowlands~\cite{rowlands} and
Zhang~\cite{mmohocc}. Their methods were limited by the choice of
maps and the kind of switching (or hopping as they call it)
mechanism. MMOHOCC of Zhang~\cite{mmohocc} uses a finite number of
arbitrarily predefined chaotic maps. They use pre-defined switching
patterns to extract points from the trajectories. We propose a
different switching mechanism, one that is chaotic. We also have the
advantage of choosing from a very large number of Robust Chaos maps.
Zhang's MMOHOCC has the problem of not having Robust Chaos for any
of their maps. The maps they use (Chebyshev and Logistic) do not
exhibit chaos for all values of the parameter. They use the fully
chaotic value of $a=4$ for the Logistic Map. However, such a method
would have the draw-back of not being fully chaotic when implemented
in hardware. It is impossible to maintain a constant value for a
parameter exactly in hardware owing to noise. Another problem they
have to worry about is the presence of {\it attracting} periodic
orbits for some values of the parameters even in the chaotic regime
(this is indicated by the windows in the bifurcation diagrams). Our
method eliminates all these problems. Although our method of using
Robust Chaos maps would also settle down to periodic orbits in
finite precision, this would not be case in an analog
implementation. It is still not known what {\it repelling} periodic
orbits would translate to in finite precision and whether this
provides any real benefit in pseudo-random number generators.
\subsubsection{The Algorithm}
\begin{figure}[!h]
\centering
\includegraphics[scale=.4]{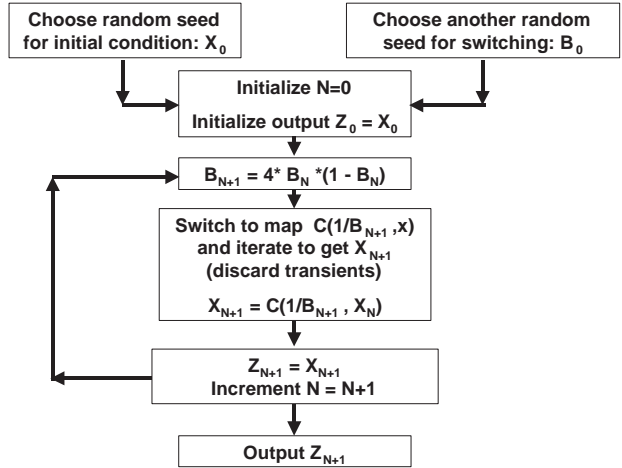}
 \caption{Flow chart for BEACH pseudo-random number generator. Here, $C(B,x)$ is any Robust Chaos family of maps. For our implementation, we have chosen $C=GL(B,x)$.} \label{fig:figflowchart}
\end{figure}

Figure~\ref{fig:figflowchart} shows the flow chart for BEACH. $\{Z_i
\}$ is the output sequence of pseudo-random numbers of length $N$.
There are two seeds $X_0$ and $B_0$ to the algorithm (numbers
between 0 and 1). $X_0$ forms the initial value of the iteration. We
assume that these seeds are generated using a {\it random} procedure
like the movement of the mouse, the speed of typing on the keyboard
or some physical characteristic (like heat dissipation) in the
hardware. In BEACH, each random number is picked from a particular
map. The maps are generated parametrically using a sequence of
$B$'s, $\{B_1,B_2,....,B_M\}$ where $M$ is the number of maps we
wish to use for switching. In our implementation, we pick one
iterate from each of the B-exponential maps ($GL(B,x)$). Thus $M=N$,
the length of the pseudo-random number sequence $\{Z_i \}$ we intend
to generate (however the $B$'s are not necessarily distinct owing to
computer precision, though in theory there are an infinite number of
them).

This sequence of $B$'s can be generated in many ways. We only need
to ensure that successive $B$'s are not sufficiently close with a
high frequency so that any two consecutive maps differ considerably.
One way of varying $B$ is by using the Logistic Map (we take 1/value
of the Logistic Map). Alternatively, $B$ can also be varied using
the standard Tent map. Such a scheme ensures that successive $B$'s
are not close to each other on the real line for most of the times.
We could also vary $B$ using the orbit of BEACH itself. For our
implementation, we use the Logistic Map for generating $B$'s.
Although varying $B$ according to the Logistic Map does not give a
random sequence of $B$'s, it is sufficient for the purpose of
switching maps. Each of the B-exponential maps are periodic because
of finite precision, but as we showed in
Section~\ref{sectionswitching}, by switching between maps, the
average period length increases considerably.

We limit the value of $B$ to 10,000 because the maps tend to the
unit function for very large $B$. This may result in small period
lengths or fixed points owing to limitations of precision on a
computer (values very near to 1 may be rounded off to 1 which
becomes 0 in the next iteration). If an iterate of the Logistic Map
is lesser than $10^{-4}$, it is replaced by an iterate of the
B-Exponential Map. If the iterate of B-Exponential Map is also less
than $10^{-4}$, then the Logistic Map iterate is set to $10^{-4}$.
Thus, we ensure that $B$ does not exceed $10,000$. The final
obtained value of $Z_N$ is between 0 and 1 in double precision. To
convert this to an integer, we multiply the iterate by $2^{52}$
(similar to Zhang's method~\cite{mmohocc}).

We also ensure that we do not use 0.75 as the seed for $B_0$ since
it is the fixed point of the Logistic Map. The other disallowed
seeds are 0 and 1 for obvious reasons.

The implementation of the algorithm was written in ANSI C in double
precision floating point arithmetic. It is very hard to analytically
determine the period of BEACH. Theoretically, robust chaos implies
that there are no stable periodic orbits and we also know that the
measure of periodic orbits is zero (in Full Chaos). However, when
implemented on a computer, all orbits are periodic owing to limited
precision. Since we have implemented BEACH in double precision
arithmetic, the number of chaotic maps available for switching is
around $10^{300}$ which would imply a substantial increase in
average period length. As we are switching in a chaotic fashion,
consecutive maps from which random numbers are extracted will be
considerably different.
\section{Randomness Evaluation of BEACH pseudo-random number generator}
\begin{figure}[!h]
\centering
\includegraphics[scale=.4]{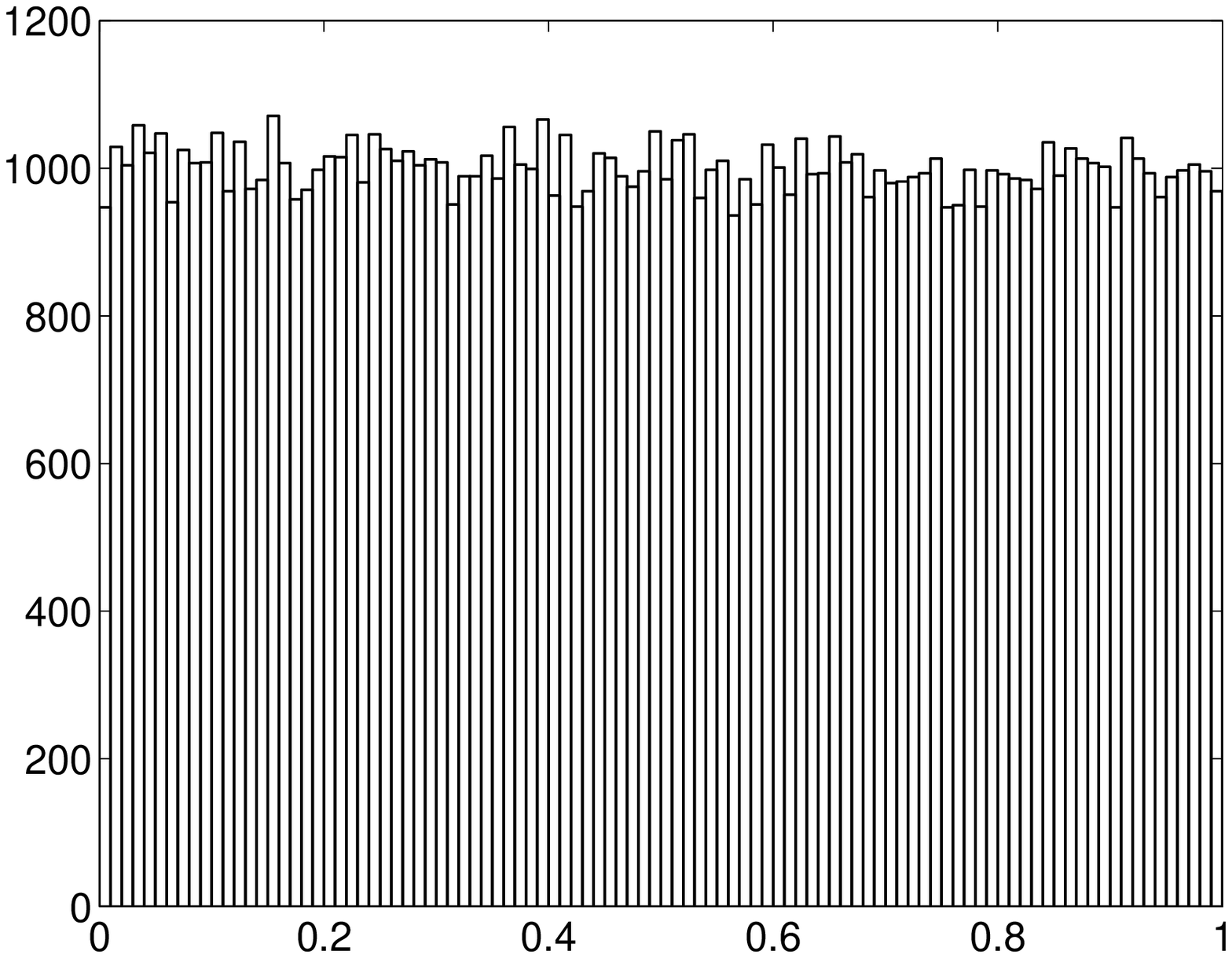}
\hspace{0.01in}
\includegraphics[scale=.4]{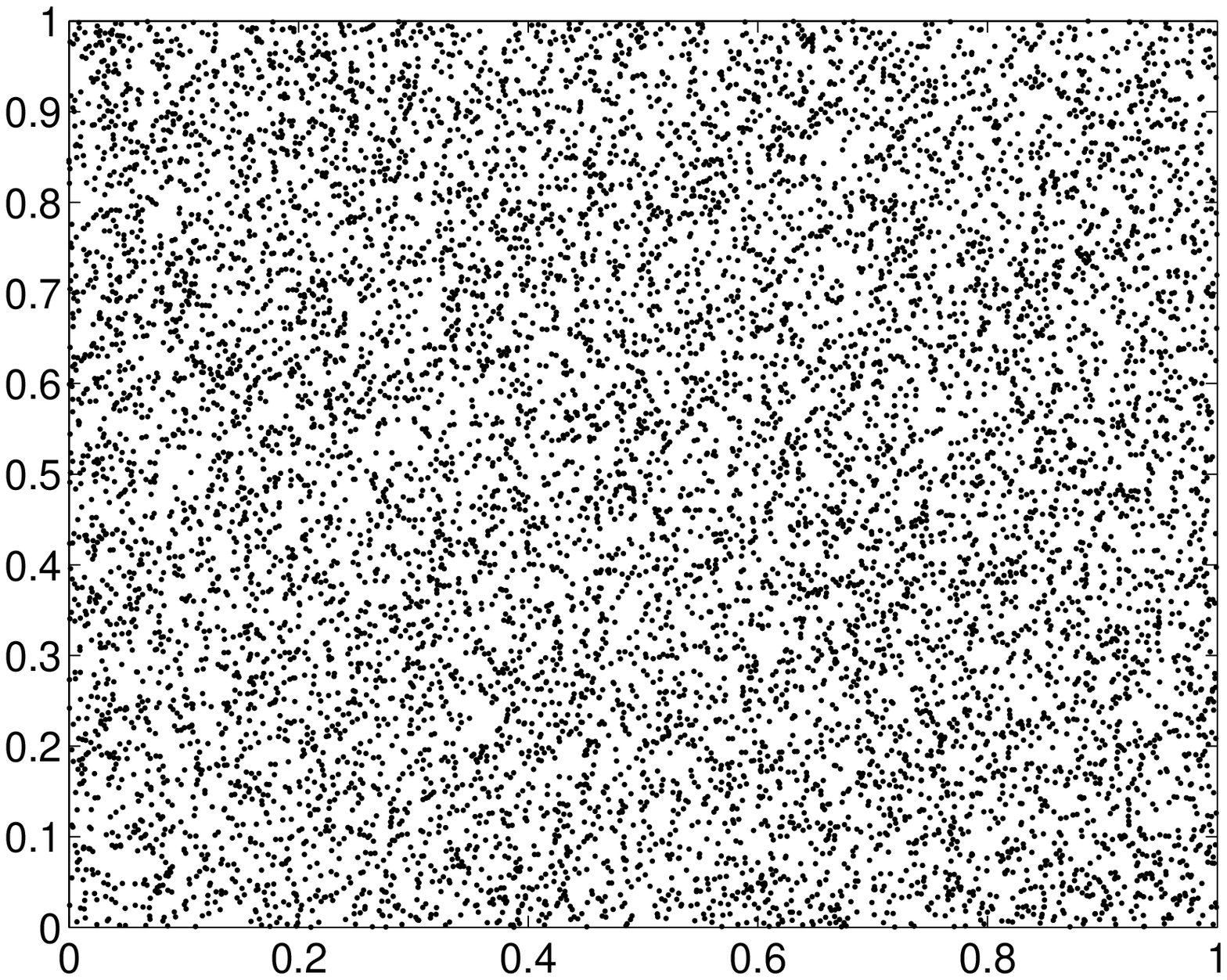}
 \caption{(a) Left: The histogram of $10^5$ pseudo-random numbers of BEACH.
 (b) Right: 2D phase space plot of $10^4$ pseudo-random values of BEACH.} \label{fig:figHist}
\end{figure}

Figure~\ref{fig:figHist}(a) shows the histogram of BEACH output
which appears uniform. The 2D phase space plots of BEACH output
shown in Figure~\ref{fig:figHist}(b) appears random. To
statistically confirm this, we tested using 3 standard test suites
-- The National Institute of Standards in Technology's Statistical
Test Suite (NIST)~\cite{nist,nist2}, George Marsaglia's Diehard
Battery of tests~\cite{diehard}, and the ENT test~\cite{ent test}.
These tests are well known in the cryptography community and are
routinely used for evaluation of random number generators. The BEACH
pseudo-random number generator successfully passed all the tests.
\subsection{Entropy, Chi-square and Mean} Shannon's
entropy is defined as $H(X)=-\sum_{x}P(x)\log_2{P(x)}$ whenever
$P(x) \neq 0$, where $P(x)$ is the probability that the random
variable $X$ is in the state $x$. Shannon's entropy is a measure of
the \emph{information density} of the data and a good measure of the
degree of disorder (randomness) in the data. We created several
binary files with the random numbers (taken as 32 bit integers) from
BEACH (0 and 1 are the two states). An optimal compression using the
ENT Pseudo-random Number Sequence Test Program (by John
Walker)~\cite{ent test}, resulted in an entropy of 1.0 bits per
symbol, consistently for all the files. Thus, the program was unable
to compress the file. This is a strong evidence that BEACH is a good
pseudo-random number generator. This is supported by the fact that
the file also passed the Lempel-Ziv Compression test which is a part
of the NIST Statistical Testing Suite.

The chi-square test is a very basic test of randomness.
Knuth~\cite{knuth taocp 2} gives a detailed treatment of the
chi-square test. The chi-square distribution is computed for a
sequence file and expressed as an absolute number and a percentage
which indicates how frequently a truly random sequence would exceed
the value calculated. This percentage is a measure of the
randomness. If the percentage is less than 1\% or greater than 99\%,
then the sequence is not random. Percentages between 90\% and 95\%
and 5\% and 10\% indicate the sequence is ``almost suspect'' to be
non-random ~\cite{knuth taocp 2}. Sequences generated by BEACH were
within 25\% to 75\% consistently.

The mean of 1 billion bit sequences of BEACH was consistently at 0.5
for 1 bit word length and 127.5 for 8 bit word length. This is
reported as part of the ENT test (Table~\ref{tab:tabent} in
Appendix). The serial correlation was also very low, of the order of
$10^{-6}$ for a billion bit sequence. In addition to this, ENT
program carried out Monte Carlo Value of Pi test. Each successive
sequence of 24 bits are used as X and Y co-ordinates within a
square. If the distance of the randomly-generated point is less than
the radius of the circle inscribed within the square, the 24-bit
sequence is considered a \emph{hit}. The percentage of hits is used
to calculate the value of $\pi$. For very large streams (this
approximation converges very slowly), the value will approach the
correct value of $\pi$ if the sequence is close to random. For
BEACH, the error percentage was consistently 0.0\% (statistically).
For the complete ENT test results, visit
$http://mahesh.shastry.googlepages.com/entres/$.
Table~\ref{tab:tabent} in Appendix lists the value of these
parameters for different lengths of generated random bits. \\
\subsection{Other well known statistical test suites} BEACH random numbers also
passed the NIST Statistical Test Suite~\cite{nist} which consists of
15 tests. \emph{Passing} of a test in NIST Suite implies a
confidence level of 99{\%}. In other words, when the p-value is more
than the passing level, the test is considered passed with a
confidence level of 99{\%}. The details of the 15 tests in the NIST
suite and their interpretation can be found in~\cite{nist}.

%
%
%

The Diehard Battery of Tests of George Marsaglia~\cite{diehard} are
collectively considered to be one of the most stringent statistical
tests for randomness. Ten streams of 1 billion bits each were
generated using ten different random seeds. Each of the seed was
chosen randomly from 10 equally spaced intervals from the set
$(0,1)$. The criteria for passing a Diehard test is that the p-value
should not be 0 or 1 up to 6 decimal places. BEACH passed all the
tests recommended in the Diehard Battery. The test results are
tabulated in Table~\ref{tab:tabdiehard} in the Appendix. The full
results of the Diehard Tests are available
\href{http://www.personal.psu.edu/mcs312}{at}
$http://mahesh.shastry.googlepages.com/diehard/$.

Furthermore, we found that BEACH successfully passes all the tests
for extremely large sequences (we tested up to 10 Gb). In general,
it is true that passing of these stringent tests only means a
failure to falsify that the sequence is random. It does not mean
that the sequence is actually random. However, since we have shown
empirically that switching of chaotic maps does considerably
increase the average period length, this may be the reason for the
success of BEACH. An open problem is the determination of the exact
relationship (on the lines of those established by Grebogi {\it
et.al.}) between the average period length, computer precision,
correlation dimensions of the maps, the number of maps being
switched and the type of switching. Such a relationship will help us
determine a bound on the average period length of BEACH
pseudo-random number generator.

\section{Conclusions}
We have investigated the effect of different switching strategies on
the average period length of simple chaotic maps (Tent map, Logistic
Map and Skewed Tent maps) when implemented in finite precision. We
have found that the average period lengths are in the following
order (from smaller to bigger): No switching $<$ Sequential
switching $<$ Chaotic switching $<$ Random switching. Furthermore,
the average period length increases with the number of maps being
switched (we did not report the actual graphs in this case for want
of space). We then introduced a generalization of the Logistic map
which exhibits Robust Chaos and developed a pseudo-random number
generator using switching of Robust Chaos maps for the first time.
Robust Chaos which exhibits no attracting periodic orbits seems to
be desirable for cryptographic applications, especially for hardware
(analog) implementations since noise induced variations of the
parameter does not result in windows. We have shown that our
proposed pseudo-random number generator successfully passes
stringent statistical tests of randomness which are well accepted in
the cryptographic community.
%
%




%
%
\section*{Acknowledgements} The authors would like to thank Dr.
Sutirth Dey of IISER, Pune, for stimulating discussions on the
B-Exponential Map. Nithin Nagaraj would like to express his sincere
gratitude to NIAS, CSIR and DST, Govt. of India, for providing with
travel grants to present this work at the International conference
on Nonlinear Dynamics and Chaos: Advances and Perspectives,
September 17-21, 2007, Aberdeen, held on the occasion of Prof. Celso
Grebogi's 60th birthday. We thank the reviewers for useful comments.

\section*{Appendix}
\begin{table}[!h]
\caption{Results of ENT on 3 bitstreams generated by BEACH.}
 \label{tab:tabent}
\begin{tabular}{|c|c|c|c|c|c|}
  \hline
  Length & Entropy & Chi-square  & Arithmetic & Monte Carlo & Serial\\
  &(per bit)&distribution(\%)& mean & value of $\pi$ (error \%)&correlation coeff
  \\
  \hline
  100 $Mb$ & 1.000000 & 50.00 & 0.5000 & 0.01 & 0.000151  \\
  500 $Mb$ & 1.000000 & 50.00 & 0.5000 & 0.00 & 0.000024 \\
  1 $Gb$   & 1.000000 & 75.00 & 0.5000 & 0.01 & 0.000035 \\
  \hline
\end{tabular}
\end{table}
\begin{table}[!h]
\centering \caption{Results (p-values of hypothesis testing) of
Diehard battery of tests on 10 bitstreams generated by BEACH, each
of length 1 Gb. The criteria for passing a Diehard test is that the
p-value should not be 0 or 1 up to 6 decimal places.}
\label{tab:tabdiehard}
\begin{tabular}{|c|c|c|c|c|c|c|c|c|c|c|c|}
  \hline
  Test & Seed & Seed  & Seed & Seed  & Seed  & Seed  & Seed  & Seed  & Seed  & Seed   \\
  No. & 1 & 2 & 3 & 4 & 5 & 6 & 7 & 8 & 9 & 10 \\
  \hline
  1 & .871  & .221 &  .250  & .922  & .014  & .154  & .069  & .050 & .790 & .515 \\
  2 & .971  & .527 &  .701  & .273  & .448  & .946  & .292  & .460 & .902 & .113 \\
  3 & .761  & .942 &  .321  & .679  & .407  & .585  & .519  & .587 & .448 & .962 \\
  4 & .374  & .801 &  .193  & .098  & .799  & .117  & .651  & .437 & .105 & .015 \\
  5 & .419  & .665 &  .317  & .324  & .040  & .876  & .678  & .507 & .468 & .075 \\
  6 & .925  & .199 &  .869  & .401  & .978  & .998  & .427  & .930 & .901 & .268 \\
  7 & .005  & .192 &  .935  & .611  & .505  & .621  & .729  & .339 & .125 & .773 \\
  8 & .516  & .549 &  .634  & .539  & .092  & .483  & .842  & .053 & .171 & .576 \\
  9 & .445  & .542 &  .080  & .964  & .724  & .773  & .807  & .136 & .383 & .806 \\
  10 &.244  & .084 &  .982  & .779  & .355  & .088  & .678  & .324 & .185 & .059 \\
  11 &.717  & .440 &  .045  & .269  & .280  & .376  & .542  & .873 & .589 & .952 \\
  12 &.566  & .702 &  .008  & .900  & .145  & .065  & .427  & .784 & .292 & .065 \\
  13 &.544  & .636 &  .096  & .376  & .768  & .922  & .324  & .903 & .987 & .677 \\
  14 &.804  & .835 &  .845  & .302  & .390  & .791  & .235  & .567 & .802 & .746 \\
  15 &.082  & .512 &  .523  & .875  & .713  & .604  & .704  & .909 & .482 & .119 \\
\hline
\end{tabular}
\\ \vspace{0.1in} The names
of the tests (1-15) are: Birthday Spacings Test, Overlapping
5-Permutation Test, Binary Rank Test for $31 \times 31$ Matrices and
$32 \times 32$ Matrices, Binary Rank Test for $6 \times 8$ Matrices,
Bitstream Test, Tests OPSO, OQSO and DNA, Count-The-1's Test On A
Stream Of Bytes, Count-The-1's Test for Specific Bytes, Parking Lot
Test, Minimum Distance Test, 3Dspheres Test, Squeeze Test,
Overlapping Sums Test, Runs Test, Craps Test.
\end{table}

\end{document}